# The influence of SARS-CoV-2 variants on national case fatality rates

William A. Barletta

*Department of Physics, Massachusetts Institute of Technology*

*Cambridge MA* 02139

## Abstract

**Background:** During 2021 several new variants of the SARS-CoV-2 virus appeared with both increased levels of transmissibility and virulence with respect to the original wild variant. The Delta (B.1.617.2) variation, first seen in India, dominates COVID-19 infections in several large countries including the United States and India. Most recently, the Lambda variant of interest with increased resistance to vaccines has spread through much of South America.

**Objective:** This research explores the degree to which new variants of concern 1) generate spikes and waves of fluctuations in the daily case fatality rates (CFR) across countries in several regions in the face of increasing levels of vaccination of national populations and 2) may increase the vulnerability of persons with certain comorbidities.

**Methods:** This study uses new, openly available, epidemiological statistics reported to the relevant national and international authorities for countries across the Americas, Europe, Africa, Asia and the Middle East. Daily CFRs and correlations of fatal COVID-19 infections with potential cofactors are computed for the first half of 2021 that has been dominated by the wide spread of several "variants of concern" as denoted by the World Health Organization.

**Results:** The analysis yields a new quantitative measure of the temporal dynamics of mortality due to SARS-CoV-2 infections in the form of variations of a proxy case fatality rate compared on a country to-country basis in the same region. It also finds minimal variation of correlation between the cofactors based on WHO data and on the average apparent case fatality rate.

Keywords: Variants of concern, case fatality rate, country correlations

## 1. Introduction

The period from mid-November, 2020 to the present has seen the rapid spread of several "variants of concern" of the SARS-CoV-2 virus [1] that have appeared across many highly populated nations with both increased levels of transmissibility and virulence. Most recently the Delta (B.1.617.2) variation has become the dominant cause of COVID-19 in several large countries including the United States and India. One question is whether the new variants increase the susceptibility for severe consequences for people with common comorbidities. Reference [2] examined that question with respect to the original stain of the virus using a cutoff date of statistics of 30 December 2020. Since then the number of reported cases of COVID-19 has risen from 82.9 million to 193.4 million. Over that same period the number of deaths rose from 1.81 million to 4.15 million. [3,4] During 2021, more than 3.7 billion people have received at least one dose of an anti-COVID vaccine and more that 1 billion are considered fully vaccinated. [5]

## 2. Objective

The starting point of the analysis is an examination of the time behavior of the pandemic averaged CFR. Figure 1 that shows the pandemic averaged CFR for the period from November 1 to the cutoff date of this study (July 18, 2021) when variants of concern in addition to Alpha were widespread.

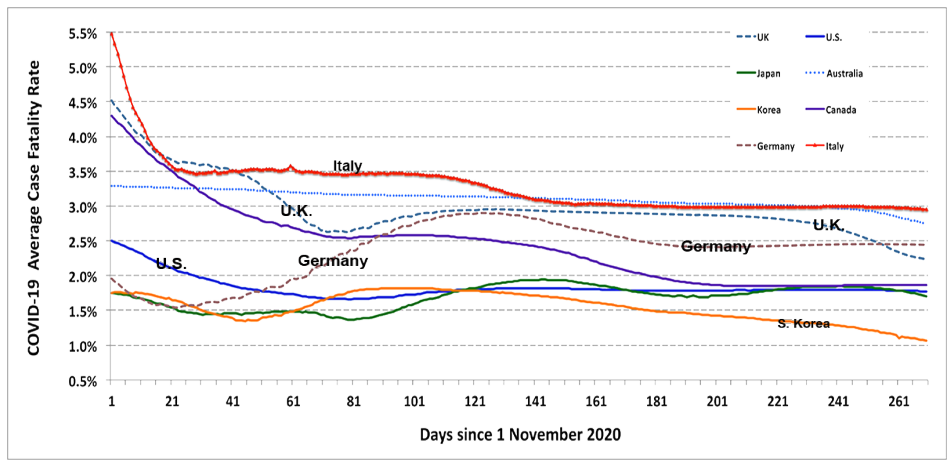

Figure 1. The pandemic averaged CFR for the period during which other variants of concern became widespread

The (red) curve for Italy emphasizes the degree to which the very high values of average CFR in the first several months of the pandemic greatly reduce the sensitivity of this measure in a search for convincing evidence of waves of increased virulence of SARS-CoV-2 variants that have spread in 2021. The figure does indicate a long period in which the CFR in the UK rose, probably due to the B.1.117 variant and then fell as the effects of Britain's vigorous vaccination and infection characterization programs took hold. Similarly, the rise in the German rate may derive from the same B.1.117 variant. However, laboratory data characterizing the nature of the infection were not available to substantiate that hypothesis. The fluctuations in the average rates for Australia, Japan, Korea, and the U.S. are not readily explainable from these pandemic averaged data.

As the pandemic averaged data are suggestive but far from dispositive, further analysis requires one to introduce a proxy measure of CFR that is more sensitive to temporal variations in virulence but is far less sensitive to the irregularities in the timing of government reports of fatalities ascribed to COVID-19 infections, an example of which appears in Figure 2. The manifest irregularities in reporting with far fewer reports on the weekends and the statistical noise necessitate introducing a proxy for estimating the daily case fatality rate.

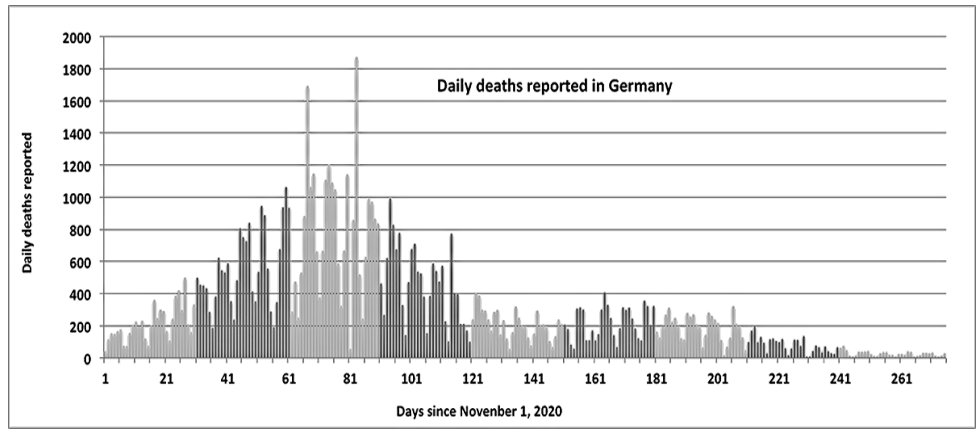

Figure 2. Daily deaths due to COVID-19 as reported in Germany. The irregularities in reporting on a weekly cycle are manifest.

Using an appropriate proxy rate, one can explore whether the daily CFR in several countries shows evidence of more virulent variants taking hold and/or whether a vigorous program of COVID-19 testing and vaccination decreases the mortality rate of the disease.

A second question is whether these new variants of concern have a different susceptibility to co-morbidities and economic factors that did the original wild variant of the SARS-CoV-2 virus. To answer this question one can analyze the correlations of potential contributing co-factors during the period of January1, 2021 to July 18, the data cutoff for this research, over the same set of countries studied in [2].

3.Method of Analysis

To explore the influence of variants of concern one introduces a credible proxy for daily case fatality rates, pCFR, which will be sensitive to the spread of variant throughout a country. The definition of pCFR and the subsequent analysis of it time distribution on a country-by-country basis are presented in Sections 4 and 4.1 respectively.

To evaluate any changes in the susceptibility to co-factors one can follow the method of reference [2], in which the input data are based only on national epidemiological statistics reported to the relevant national and international authorities. For consistency of analysis, this study considers the same sample of countries as reference [2]. The regions and their constituent countries are given in the Appendix.

**4. Results**

The analysis of this paper evaluates 1) the linear correlation between a proxy CFR for country pairs and 2) the linear correlation of the pandemic averaged CFR and potential cofactors using the usual Pearson "product moment correlation" of Equation (1).

$$r = \frac{\sum (x_i - x_{average})(y_i - y_{average})}{\sqrt{\sum (x_i - x_{average})^2 \cdot \sum (y_i - y_{average})^2}} \qquad (1)$$

The resulting values of the correlations show minimal differences between those with a data cutoff of December 30, 2020 and the data with a cutoff of July19, 2021. The one exception is the jump by a factor >2 in the correlation between the apparent CFR and number of COVID-19 cases per 1 million persons. That jump persons may be due to the increased virulence of the variants of the concern that have spread during 2021 especially in countries without robust vaccination programs. Doubtless they are also affected by the pervasiveness of national vaccination programs.

To explore this question and compare experience in several countries one can look at the time series of levels of daily case fatality rates. Unfortunately, as shown in Figure 2, actual data will be very noisy, subject to uneven reporting of both new cases and deaths attributed to COVID-19 as well as to significant statistical variations in the daily data.

To mitigate these deficiencies in medically definitive data, one can introduces pCFR as a plausible proxy for the apparent daily case fatality ratio. The model then overlays those data with a rolling 14-day average of the results to suggest the actual temporal trends in the virulence of SARS-CoV-2 infections. Noting that grave consequences of infection often appear within a 7-day period from 14 to 21 days after reported infection, one can define the proxy pCFR by Equation (2). The results of the analysis depend only weakly on the period over which pCFR is computed.

$$pCFR = \frac{\text{New Deaths on day N}}{[\text{Total cases on day (N-14) - Total cases on day (N-21)}]/7} \qquad (2)$$

One could equally well average the infection rate over the period from 28 to 14 days. Figure 3, which displays the time sequence of the pCFR for the United States, shows only weak evidence for differential rates of mortality in February and March 2021 due to the B.1.117 variant even though that variant appeared in the U.S. in January 2021 [7], before any noticeable fraction of the population had received vaccinations. In Ref. [7], the CDC had predicted a peaking of infections due to B.1.117 in March, 2021. That might account for the slight increase in pCFR seen in March in the U.S.

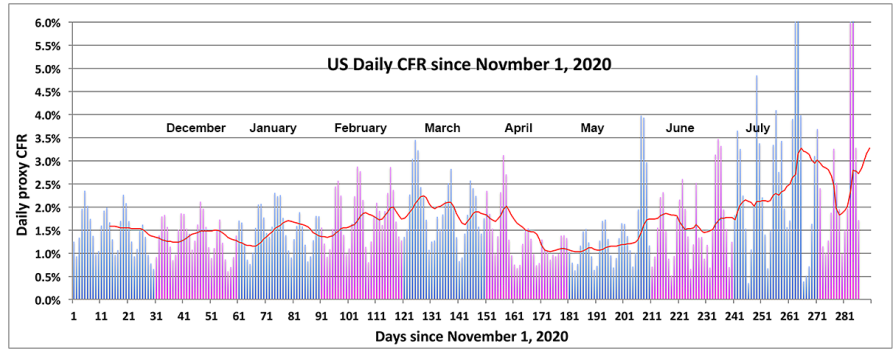

Figure 3. Daily pCFR in the United States since November 2020. Months are in light and dark bands. The dark line is a 7-day rolling average.

In contrast the U.K. reported infections from the B.1.117 (Alpha) variant in mid-December. This variant spread quickly raising the pCFR to about 3% (based on a 14-day rolling average) before a significant fraction of the U.K. population could be vaccinated. Figure 4 shows a marked increase in the pCFR in late December and January. By March ~30% of the U.K. population had received their vaccinations [8], and by April over 80% of the population ≥59 years of age had been vaccinated. The pCFR began a steady decrease in March. It is probable that thanks to a vigorous program of both testing for SARS-CoV-2 infection (at twice the rate of the U.S.) plus vaccination with more than 60% of the total U.K. population immunized by May 1, 2021, the pCFR in the U.K. is now well below 0.5%, lower than any country in Europe.

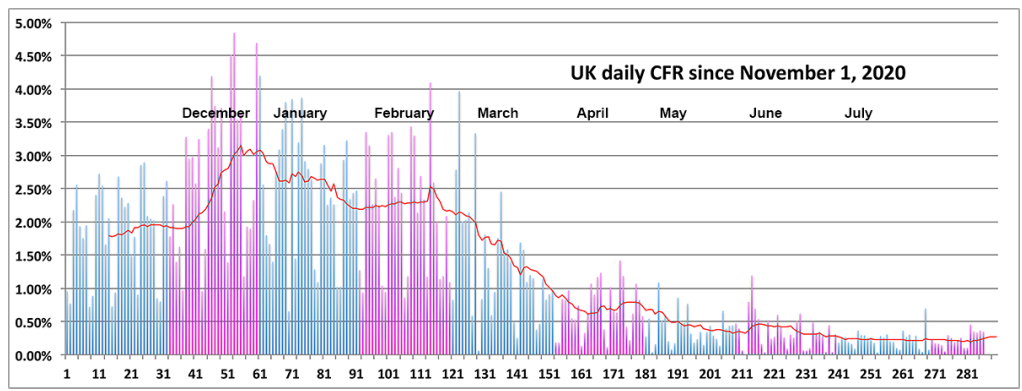

Figure 4. Daily pCFR in the United Kingdom since November, 2020. Months are in light and dark bands. The dark line is a 7-day rolling average.

Figure 5 illustrates the case in Germany—intermediate between that of the United Kingdom and the United States. The slower spread of the B.1.117 variant probably explains the increase of pCFR during January and March that parallels the rise in pCFR in Britain. This behavior offers further evidence that the B.1.117 variant is more virulent than the original wild strain of SARS-CoV-2 from Wuhan. Probably due to the excellent German health care system, in mid-March the pCFR began to decrease toward its pre-B.1.117 level. Yet as vaccinations and testing for COVID-19 in Germany have lagged well behind the levels in the U.K., reaching 10% full vaccination only in early May, 2021, one observes pCFR increasing by the end of May, likely due to the more virulent B.1.617.2 (Delta) variant. That behavior is similar to what has been seen in the United States (Figure 3). By the beginning of July, full vaccination in Germany had reached 40%, and the pCFR showed signs of lessening to approximately 3%.

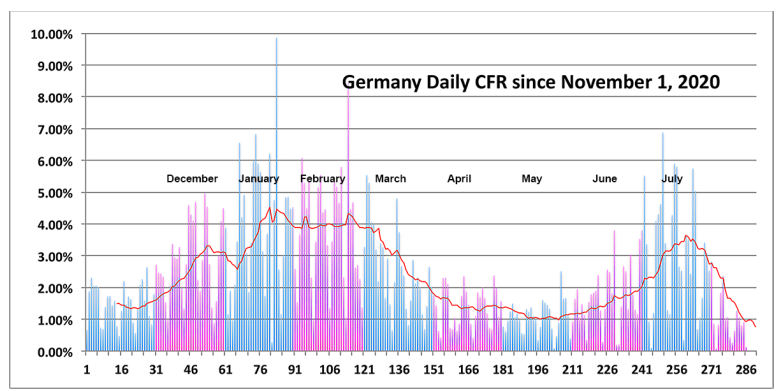

Figure 5. Daily pCFR for Germany since November 2020. Months are in light and dark bands. The dark line is a 7-day rolling average.

**4.1 The variant dominated period**

An objection to the initial analysis of the previous section, is that the pandemic averaged case fatality ratios are dominated by the very high mortalities at the outset of the pandemic before adequate and appropriate isolation of the infected and modalities of treatment were understood. To mitigate that objection, one can focus only on the period of November 2020 through July 2021 that is dominated by surges of variants of concern that were described at the time of the WHO's designation to have higher transmissibility and perhaps higher virulence than the original wild strain of SARS-CoV-2.

The statistics with a cutoff of July 18, 2021 yield Figure 6. The red dashed curve accounts for the correlations only during the variant-dominated period. The solid black and grey curves indicate the correlations throughout the pandemic computed from data with the indicated cutoff dates and using the pandemic averaged CFR. Perhaps remarkably, with one exception one sees no great differences between the variant-dominated and the pandemic averaged curves. The doubling of the correlation between the average CFR with total number of COVID-19 deaths per capita is likely due to deaths in unvaccinated populations caused by the B.1.617.2 (Delta) variant, which now accounts for 80% of infections in the U.S. [9]

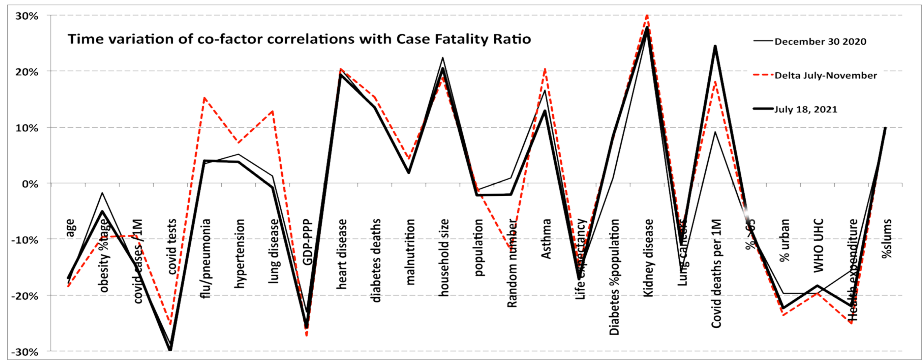

Figure 6. Co-factor correlations in the variant-dominated period (red dashed curve) compared with values of pandemic averaged CFRs.

A potential source of misinterpretation of the global statistics is the variation of the correlations of cofactors from one region to another as well as from the overall global value. Figure 7 compares the correlations of the average case fatality rate for six commonly cited cofactors for the period dominated by the variants of concern.

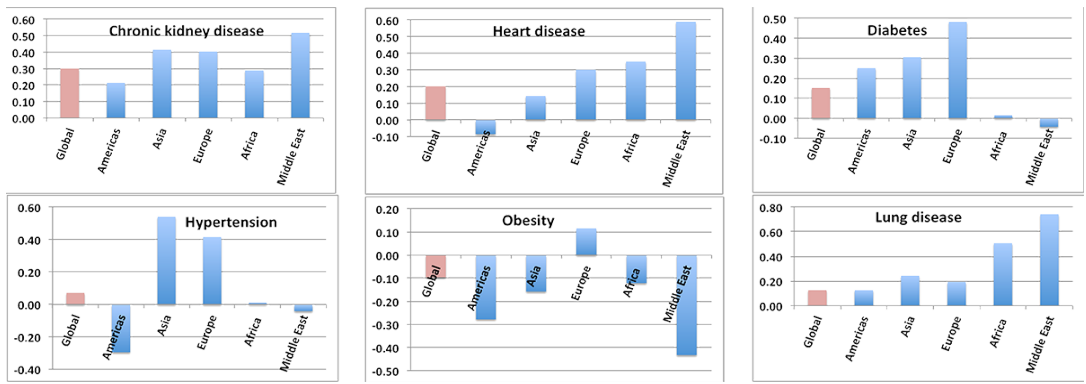

Figure 7. Variation of correlations with the average CFR over the period from November 1, 2020 to July 18, 2021

An examination of statistics [4] from Israel, which instituted a vigorous vaccination program early in 2021 might have shed light on the role of testing and vaccination to suppress the serious consequences of infections with SARS-CoV-2. As seen in Figure 8, while the smoothed data [4] do show evidence of an increase in pCFR during late May and June consistent with spread of the Delta variant in Israel, they are too noisy to allow firm conclusions. Moreover, across all elements of the population of Israel, the overall vaccination rate is only 60.1%. However, the rate for persons ≥60 years of age is 80%.

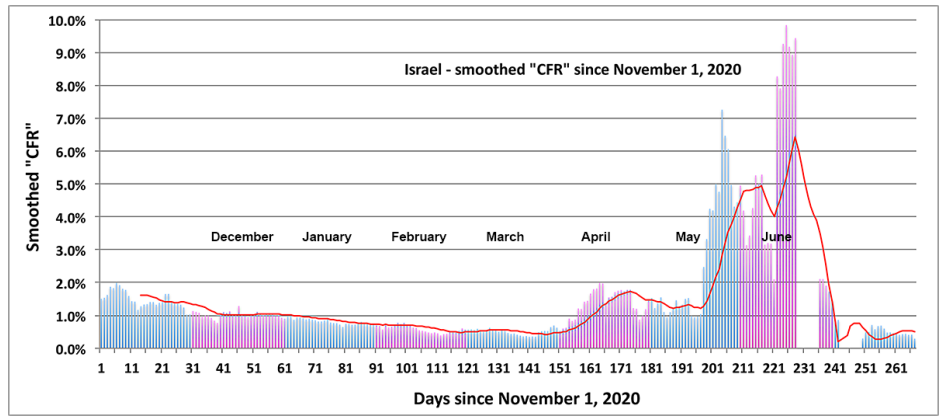

Figure 8. Smoothed daily "CFR" in Israel from November 1, 2020. The dark line is a 14-day rolling average

An examination of the smoothed data [4] from India in a manner similar to that used to generate Figures 2 through 4 provides a useful day-by-day comparison [4] among the countries. As it has been in the United States in July 2021, the dominant variant in India is B.1.617.2 that started to become common in India in March, 2021. [10] Although the Indian data has less statistical noise than the data from Israel, the smoothed day-by-day statistics, presented in Figure 8, allow for a clearer look at temporal trends than do the raw data. Consistent with the increasing pervasiveness of the B.1.617.2 variant, pCFR displays a significant increase from its February low of less than 1% rising to 2% by late July. The rolling 14-day average (dark line) shows a clear increase in pCFR as the B.1.617.2 variant has become ever more pervasive.

The results of showed in Figure 9 are likely to include little effect of India's program of vaccination that uses five different vaccines. By mid-July, 2021, only 5% of its population had been vaccinated. [11] Moreover, all the vaccines evaluated against B.1.617.2 appear to be roughly 10% less effective (at the 95% confidence level) in controlling the development of COVID-19 in patients with the B.1.617.2 variant. [12].

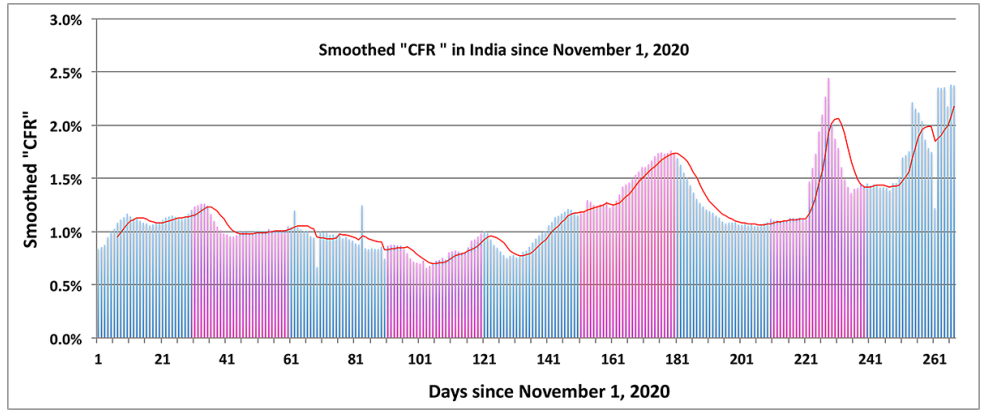

Figure 9. The smoothed values of the CFR proxy in India since November 1, 2020. The dark line is a 7-day rolling average

A recently designated variant of interest is C.37 (Lambda) which has been circulating widely in South America, having first been seen in Peru in July, 2020 [13, 14]. As shown in Figure 10, Peru itself saw a strong spike in pCFR in January and February, 2021, reaching 20%. Since then pCFR has dropped gradually to roughly 5%. By July, 2021 the rate of full vaccination of the Peruvian population had reached only slightly more than 10%. [4] In addition, in analysis of the C.37 virus, Kimura et al. [14] identified a spike structure that accounts for higher resistance to vaccine induced immunity than displayed by the original wild variant. Hence, the commencing of a vaccination program in Peru cannot by itself account for the decline in the proxy case fatality rate.

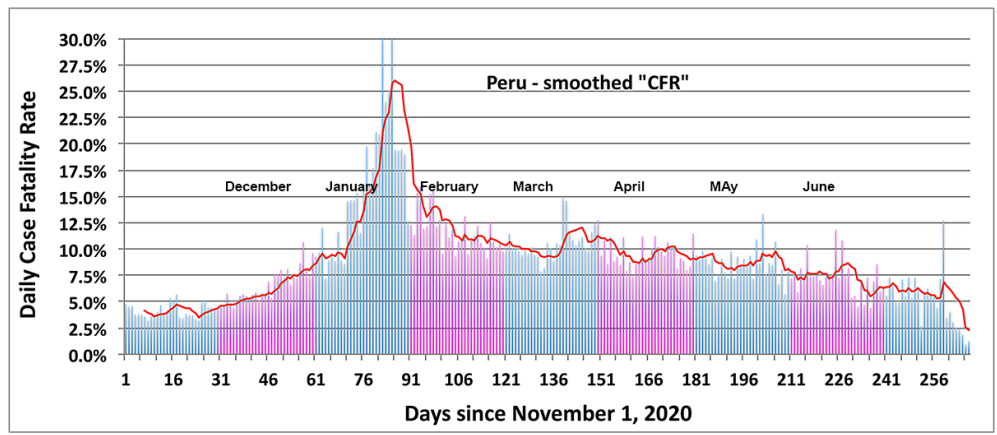

Figure 10. The behavior of the proxy CFR in Peru since November 2020 based on the smoothed data of Ref. [4]. The dark line is a 7-day rolling average

The South American scene is complicated by the simultaneous circulation of multiple variants of concern, particularly in its most populous country, Brazil. Figure 11 provides an example for Columbia.

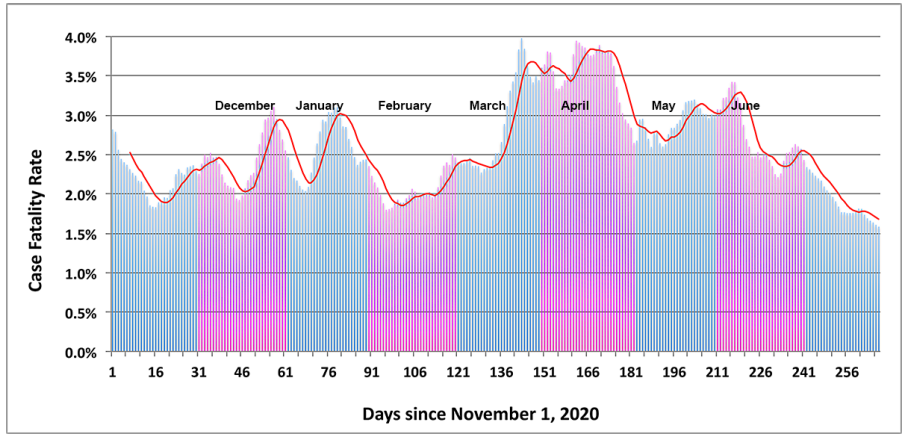

Figure 11. The proxy CFR, pCFR, for Columbia since November, 2020 based on the smoothed data of Ref. [4]. The dark line is a 7-day rolling average.

**4.2 Correlations between regional case fatality rates**

Comparison between plots of the temporal behavior of the proxy CFR (Figures 2 – 4 and Figures 8 – 11) can be made qualitative by computing the *r*-value for pairs of countries grouped into regions. One such set of calculations is displayed in Figure 12. The uncertainty in the correlation values is approximately ±0.05.

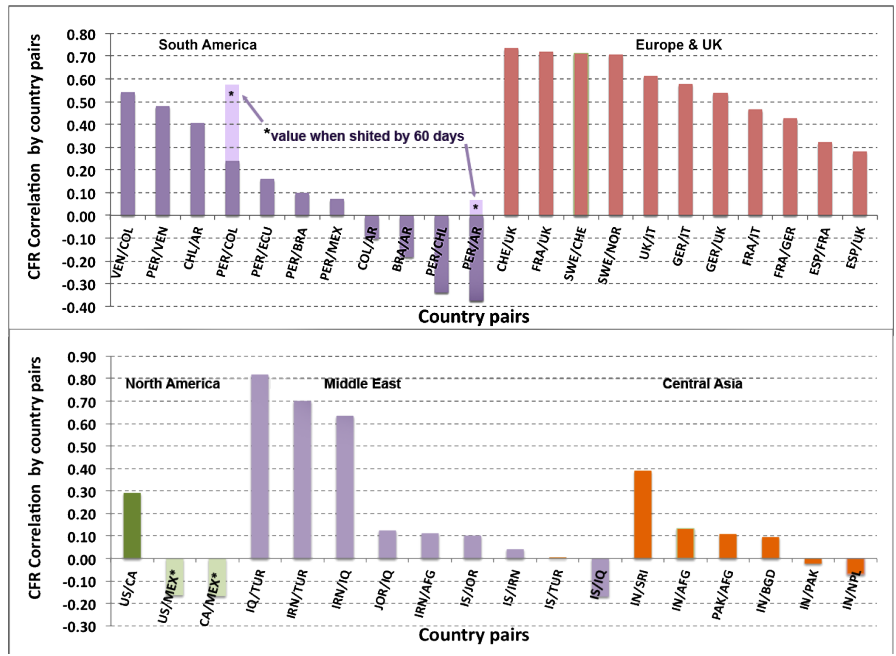

Figure 12. Correlations of pCFR between pairs of countries in five regions

The figure displays a strong to moderate correlation between countries in which travel was possible during the pandemic the spread of the B.1.117 variant from the UK. The country pairs in the Middle East show that effect strongly. The negative correlations simply indicate very different courses of infection, treatment modality, and vaccination between the country pairs. No underlying causative factors are apparent. An example of such a negative correlation is given in Figure 13, which compares the time variation of pCHR in Peru and in Argentina.

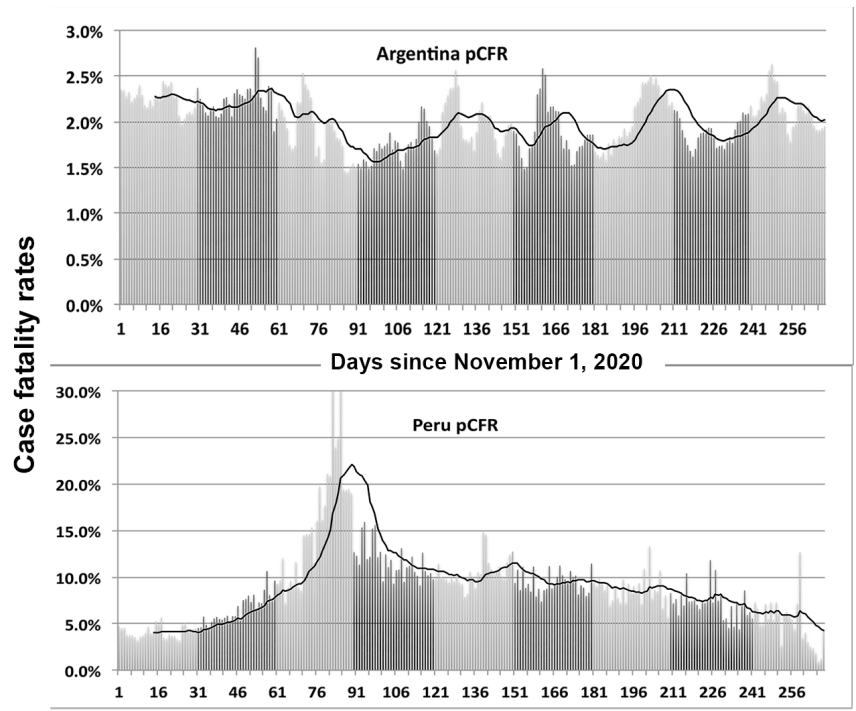

Figure 13. Example of a country pair exhibiting a significant negative correlation. The dark curves represent a 14-day rolling average

The distribution shows no indication of the Lambda variant moving from Peru to and through Argentina. Moreover, one sees no evidence of the Delta variant causing a spike in mortality in either Argentina or Peru. In fact, the correlation of the distribution of new cases in these two countries is only 0.14. However, shifting the Peruvian distribution 60 days later increase the correlation of new cases to 0.86. Moreover, the Delta variant has only been recorded in Argentina very recently [15]. The wave in new cases in Argentina could be caused differences in effective treatment in the two countries or by differences in predominant variants of SARS-CoV-2.

In contrast with the case of Argentina, the light purple bar over the PER/COL pairing shows that shifting the Peruvian profile 60 days later in time, introduces a much larger similarity with the Columbian profile. Also, the correlation of new cases shifts to 0.68. One may interpret these results as indicating the amount of time needed for the Lambda variant to spread widely into Columbia, where its prevalence is now high. [14]

For North America one observes only a moderate correlation in pCFR between the US and Canada. That low value may be explained differences in the US and Canadian health care systems and by the fact that the border between the countries has been closed since March 2020. (The US/Canada case correlation is 0.57.) Unfortunately, the calculations of pCHR for US/Mexico and Canada/Mexico are unreliable because of Mexico's large adjustments in the number of fatalities attributed to COVID-19 during two distinct periods. Those adjustments do not effect the US/Mexico correlation in new case that is 0.72, while the Canada/Mexico correlation is 0.12.

5. Discussion and Conclusions

The proxy for the daily case fatality rate, pCFR, computed over a smoothed distribution of deaths attributed to COVID-19 infections provides a useful metric to track the national dynamics of the spread of SARS-CoV-2 variants of concern overlaid with that country's vaccination implementation program. The example of the United Kingdom is instructive in this regard. A clear arise in fatality rate due to the increased virulence of the B.1.117 variant is followed by the sharp decline in the daily case fatality rate to about 0.25% thanks to the UK's aggressive program of vaccination. That low rate has persisted despite the spread of the Delta variant through the UK.

The increases in daily pCFR in the US and Germany correspond in timing with the rapid spread of the Delta variant and support the characterization of Delta as being more virulent than the original wild strain of SARS-CoV-2. The pCFR increased in the US to about 2% by the end of July 2021 and has continued to increase in the first half of August despite the moderate success of the vaccination program in the US. In contrast with the U.S., the pCFR in Germany has decreased by early August to a value of roughly 1%.

With respect to the spread of the B1.117 variant the temporal profile of pCFR for the US is too variable and the data set is too noisy to confirm the timing of the spread of that variant through the US. In Germany, however, the temporal distribution of the proxy CFR of Figure 5 shows a distinct peak with a timing that coincides with the spread of B1.117 through Europe.

The apparent increases in the virulence of the *variants of concern* might be due to increased susceptibility to severe infection in persons with certain comorbidities. Following the method of Ref. [2] for a data set limited to November, 2020 through mid-July, 2021, one finds minimal quantitative differences with the conclusion of Ref. [2] that most commonly cited co-morbidities do not *in and of themselves* increase the risk of serious (and possibly fatal) consequences of SARS-CoV-2 infections. However, one cannot ignore that many persons with such conditions often suffer from multiple co-factors and have a depressed level of immune function that can worsen the effects of a COVID-19 infection. The prevalence of multiple co-existing conditions may also vary from region to regions, partially explaining the regional variations seen in Figure 7.

**Acknowledgements**

The author acknowledges his colleagues in the World Federation of Scientists for their encouragement to continue, expand, and report this research. The author's work has been completely self-supported, without any outside funding or other material support.

**Conflicts of Interest**

None declared.

**Abbreviations**

CFR –      Case Fatality Rate

pCFR –     Proxy case fatality rate

WHO –      World Health Organization

**References**

[1] "SARS-CoV-2 Variant Classifications and Definitions," https://www.cdc.gov/coronavirus/2019-ncov/variants/variant-info.html#Concern

[2] W.A. Barletta, "Risk Factors of SARS-CoV-2 Infections: A Global Epidemiological Study," JMIRx Med (forthcoming). doi:10.2196/28843

[3] *Worldometer*, https://www.worldometers.info/

[4] "*Our World in Data* COVID-19 Dataset," https://github.com/owid/covid-19-data/tree/master/public/data

[5] "Coronavirus (COVID-19) Vaccinations," Our World in Data, https://ourworldindata.org/covid-vaccinations, July 27,2021

[6] "World Health Rankings," https://www.worldlifeexpectancy.com/world-health-rankings based on WHO data of 2018

[7] "Emergence of SARS-CoV-2 B.1.1.7 Lineage — United States, December 29, 2020–January 12, 2021," January 22, 2021, https://www.cdc.gov/mmwr/volumes/70/wr/mm7003e2.htm

[8] Vaccinations in the United Kingdom, UK government dashboard, https://coronavirus.data.gov.uk/details/vaccinations

[9] "The Delta Variant: What Scientists Know, " New York Times, June 22, 2021, https://www.nytimes.com/2021/06/22/health/delta-variant-covid.html

[10] "A coronavirus variant first found in India is now officially a 'variant of concern,' the W.H.O. said," New York Times, May 10, 2021, https://www.nytimes.com/2021/05/10/world/asia/india-covid-virus-variant.html

[11] "India vaccination: Six months on, India's vaccine drive is lagging," BBC News, July 16, 2021, https://www.bbc.com/news/world-asia-india-56345591

[12] L. Bernal, N. Andrews, C. Gower, E. Gallagher, R. Simmons, S. Thelwall, et al., "Effectiveness of Covid-19 Vaccines against the B.1.617.2 (Delta) Variant," NEJM, DOI: 10.1056/NEJMoa2108891, July 21, 2021

[13} M. L. Acevedo, L. Alonso-Palomares, A. Bustamante, A. Gaggero, F. Paredes, C. P. Cortés, et al., "Infectivity and immune escape of the new SARS-CoV-2 variant of interest Lambda, " medRxiv, July 1, 2021, doi: https://doi.org/10.1101/2021.06.28.21259673

[14] I. Kimura, Y. Kosugi, J. Wu, D. Yamasoba, E. P. Butlertanaka, Y. L Tanaka, et al. "SARS-CoV-2 Lambda variant exhibits higher infectivity and immune resistance," bioRxiv, July, 28, 2021, doi: https://doi.org/10.1101/2021.07.28.454085

[15] "Argentina reports first local Delta variant cases in Buenos Aires," August 2, 2021, http://outbreaknewstoday.com/argentina-reports-first-local-delta-variant-cases-in-buenos-aires-17322/

# Appendix

Table A. Countries sampled in this study grouped into 5 regions are the same as in reference [2].

| Region | Population (million) | Countries |
|---|---|---|
| Americas | 977 | Argentina, Bolivia, Brazil, Canada, Chile, Columbia, Costa Rica, Dominican Republic, Ecuador, El Salvador, Guatemala, Honduras, Mexico, Panama, Paraguay, Peru, United States, Venezuela |
| Asia | 2504 | Australia, Bangladesh, China, India, Indonesia, Japan, Kazakhstan, Kyrgyzstan, Korea, Malaysia, Nepal, New Zealand, Pakistan, Philippines, Singapore, Thailand, Taiwan |
| Europe | 725 | Albania, Armenia, Austria, Azerbaijan, Belarus, Belgium, Bosnia, Bulgaria, Croatia, Czechia, Denmark, Estonia, Finland, France, Germany, Greece, Hungary, Ireland, Italy, Macedonia, Moldova, Netherlands, Norway, Poland, Portugal, Romania, Russia, Serbia, Spain, Sweden, Switzerland, Ukraine, United Kingdom |
| Africa | 768 | Algeria, Cameroon, Congo, Ethiopia, Ghana, Ivory Coast, Kenya, Libya, Madagascar, Mali, Morocco, Nigeria, South Africa, Sudan, Uganda, Zambia |
| Middle East | 487 | Afghanistan, Bahrain, Egypt, Iran, Iraq, Israel, Lebanon, Kuwait, Oman, Qatar, Saudi Arabia, Turkey, United Arab Emirates, Uzbekistan, Yemen |